\documentclass[12pt]{article}
\usepackage[utf8]{inputenc}
\usepackage{graphicx}
\usepackage{natbib}
\usepackage{color}
\usepackage{amsmath}
\usepackage{amsfonts}
\usepackage{hyperref}
\usepackage[algo2e]{algorithm2e} 
\usepackage{algorithm}
\usepackage{verbatim}
\usepackage{caption}
\usepackage{setspace}
\usepackage{subcaption}
\usepackage{multirow}
\usepackage[left=1in,right=1in,top=1in,bottom=1in]{geometry}
\newcommand{\bc}{\boldsymbol c}

\newcommand{\bpi}{\boldsymbol \pi}
\newcommand{\btheta}{\boldsymbol \theta}
\newcommand{\bA}{{\bf A}}
\newcommand{\bB}{{\bf B}}

\title{Hypothesis testing for community structure in temporal networks using e-values}

\author{Eric Yanchenko\footnote{Human and AI Center, Akita International University. Okutsubakidai-193-2 Yuwatsubakigawa, Akita, Japan, 010-1211. \url{eyanchenko@aiu.ac.jp}},  Jonathan P. Williams\footnote{Department of Statistics, North Carolina State University} and Ryan Martin$^{\dagger}$}

\begin{document}

\maketitle
\thispagestyle{empty}

\begin{abstract}
\noindent
Community structure in networks naturally arises in various applications. But while the topic has received significant attention for static networks, the literature on community structure in temporally evolving networks is more scarce. In particular, there are currently no statistical methods available to test for the presence of community structure in a sequence of networks evolving over time. In this work, we propose a simple yet powerful test using e-values, an alternative to p-values that is more flexible in certain ways. Specifically, an e-value framework retains valid testing properties even after combining dependent information, a relevant feature in the context of testing temporal networks. We apply the proposed test to synthetic and real-world networks, demonstrating various features inherited from the e-value formulation and exposing some of the inherent difficulties of testing on temporal networks.
\end{abstract}


\clearpage

\section{Introduction}
This paper focuses on testing the null hypothesis of no community structure in a temporally evolving sequence of networks \citep{holme2012temporal}. For the case of a static network, there has been considerable work on this topic \citep[e.g.,][]{lancichinetti2010statistical, bickel2016:aa, li2020aa, yuan2022testing, yanchenko2024generalized}. The key challenged addressed in the aforementioned references is selecting a sensible null model to represent ``no community structure,'' as well as deriving the null distribution of the test statistic. Indeed, even defining what it means for a network to exhibit community structure is not universally agreed upon \citep{cazabet2023challenges}. For the null model, many methods use the Erdos-Renyi \citep[][henceforth ER]{erdos1959} model \citep[e.g.,][]{bickel2016:aa} while others use the configuration or Chung-Lu model \citep{chung2002average, yanchenko2024generalized}. For deriving the null distribution, an asymptotic distribution can be found in some situations \citep{bickel2016:aa}, but when this is not possible, a bootstrap approach is also available \citep{yanchenko2024generalized}.

Despite this extensive work on static networks, to the authors' knowledge, there are no existing tests for community structure on temporal networks. Users may be tempted to simply aggregate the temporal network into a static network and then use an existing static hypothesis test for community structure. Not only does this remove any temporal variation, but it has also proven to be a poor approach for other network tasks, e.g., Influence Maximization \citep{erkol2020influence, yanchenko2024influence}. Perhaps the closest related work comes from network monitoring and surveillance \citep{woodall2017overview, jeske2018statistical}. For example, \cite{wilson2019modeling} study change-point detection in a temporally evolving network using the degree corrected block model \citep[][henceforth DCBM]{karrer2011stochastic}. Those authors assume that network snapshots are being generated independently and identically distributed until some unknown time $t^*$, at which point the data-generating process changes. The goal is to find this change-point time $t^*$. In another work, \cite{wilson2017community} derive a hypothesis test for community structure in multilayer networks, for which temporal networks could be considered as a special case. Their method identifies densely-connected nodes to compute a significance score which is then compared against the configuration model \citep{newman2006modularity}.

When testing on temporal networks, new challenges emerge in addition to inheriting those of the static setting. As in static network testing, carefully defining community structure in a temporal setting is quite difficult \citep{cazabet2023challenges}. Unique to temporal networks on the other hand, sequential realizations induce highly non-trivial dependence, making it difficult to combine results across observations using e.g., p-values.

To overcome these challenges, we develop a community detection hypothesis test for temporal networks using an e-value framework. E-values have recently been gaining popularity in testing settings \citep[e.g.,][]{vovk2021values, xu2024online}. The name comes from ``expectation'' in that their key property is their expectation is less than or equal to 1 when the null hypothesis is true. Thus, large e-values correspond to evidence against the null. Moreover, key properties of e-values address the difficulty of combining results with arbitrary dependence. As many in the network science community may be unfamiliar with e-values, we devote a sizable portion the manuscript to discussing their basic features. 

The proposed test first calculates a p-value for (static) community structure on each temporal snapshot before converting this to an e-value. The e-values are then averaged to yield an overall measure of evidence against the null hypothesis of no community structure across networks. Large e-values indicate the likely presence of community structure. By using e-values, we can easily combine information from different snapshots while still maintaining type I error guarantees even with arbitrary dependence in the network process. Additionally, the general nature of our method means it can be applied to a wide-range of data-generating mechanisms and static hypothesis tests. Indeed, we apply the method to two different static hypothesis tests as well as three data-generating mechanisms. Moreover, our experiments on real-data highlight some of the challenges of defining and testing for temporal community structure \citep{cazabet2023challenges}.

The remainder of the paper is organized as follows.  Section~\ref{sec:back} provides the necessary preliminaries before describing the details of the hypothesis testing framework in Section~\ref{sec:method}. The method is applied to both synthetic and real-world networks in Sections~\ref{sec:sim} and \ref{sec:real}, respectively. Finally, we close with an extended discussion in Section~\ref{sec:disc}.

\section{Background}\label{sec:back}

\subsection{Notation}
Let $\mathcal G=(\mathcal G^{(1)},\dots,\mathcal G^{(T)})$ be a temporal network where $\mathcal G^{(t)}=(\mathcal V^{(t)},\mathcal E^{(t)})$ is the ``snapshot'' of the network at time $t$, with node set $\mathcal V^{(t)}$ and edge set $ \mathcal E^{(t)}$, and $T$ is the number of snapshots. At time $t$, we assume that there are $| \mathcal V^{(t)}|=n_t$ nodes and $| \mathcal E^{(t)}|=m_t$ edges. In this work, we assume that all edges are undirected, but the ideas can easily be extended to directed networks. Notice that both the number of nodes and edges can change over time, but in this work, we primarily focus on the setting where the number of nodes is constant, i.e., $n_t\equiv n$ for all $t$. We also define $\bA=(\bA^{(1)},\dots,\bA^{(T)})$ as the adjacency matrix corresponding to $\mathcal G$ where $A_{ij}^{(t)}=1$ if nodes $i$ and $j$ have an edge at time $t$, and 0 otherwise. For now, we remain agnostic as to how the network is generated and evolves, but assume that there are no self-loops, i.e., $A_{ii}^{(t)}=0$ for all $i,t$. Lastly, we define $\mathcal M_1$ as a set of probability distributions over sample space $\Omega$ and $\sigma$-algebra $\mathcal F$, and $\mathcal P_0\subseteq \mathcal M_1$ as the distributions corresponding to some null hypothesis. We let $\mathbb P\in\mathcal M_1$ represent a single distribution.

\subsection{Challenges in temporal network testing}

Hypothesis testing on networks is difficult due to the inherent dependence in the data, and it is made even more difficult when working with temporal networks. Specifically, we need to combine information with unknown and non-trivial dependence. Assume we observe a temporal network $\mathcal G=(\mathcal G^{(1)},\dots,\mathcal G^{(T)})$, consisting of $T$ ``snapshots'' or realizations of the network-generating process. Each snapshot, $\mathcal G^{(t)}$, yields some information about the presence/absence of a community structure in the under-lying data-generating process. We may assume that there are existing methods to extract the evidence of a community structure for each individual snapshot, i.e., using a static hypothesis test for community structure \citep[e.g.,][]{bickel2016:aa, yanchenko2024generalized}. But how can we combine the evidence from each snapshot into an overall measure of evidence of community structure in the temporal network?
Indeed, the snapshots likely have a highly non-trivial dependence, greatly complicating their combination. If we use traditional p-values, then in general it is quite challenging to combine this information, especially with unspecified dependence \citep[e.g.,][]{benjamini1995controlling}. Therefore, we seek a method that can easily combine information about the underlying data generating structure even with unknown dependence.

\subsection{E-values}\label{sec:evals}
\subsubsection{Basics}
Before presenting the proposed test, we give some important background on e-values and their basic properties. Recently, e-values have become a popular approach for hypothesis testing \citep{wasserman2020universal, ramdas2020admissible, dey2024, xu2024online, grunwald2024safe}. Among other things, they allow for: combination under arbitrary dependence, anytime-valid stopping rules and optional continuation of experiments, but we primarily focus on the combination property. We emphasize that the results in this sub-section are not novel; this discussion is intended to provide the foundation for understanding our proposed method as described in Section \ref{sec:method}.

Before introducing e-values, we briefly review the basics of p-values.
Recall that $\mathcal M_1$ is the set of probability distributions and $\mathcal P_0\subseteq\mathcal M_1$ are the probability distributions corresponding to some null hypothesis. Then $P$ is a p-value for $\mathcal P_0$ if 
\begin{equation}\label{eq:pval}
\sup_{\mathbb{P} \in \mathcal{P}_0} \mathbb P(P\leq\alpha)\leq \alpha, \quad \alpha \in [0,1].
\end{equation}
In words, this condition requires that, for all $\mathbb P\in\mathcal P_0$, $P$ is stochastically no smaller than a uniform random variable on $[0,1]$. On the other hand, an e-value, $E$, is a $[0,\infty]$-valued random variable such that its expectation under the null hypothesis is at most 1, i.e., 
\begin{equation}\label{eq:eval}
\sup_{\mathbb{P} \in \mathcal{P}_0} \mathbb E_{\mathbb P} (E)\leq 1,
\end{equation}
where $\mathbb E_\mathbb P$ represents the expectation taken with respect to $\mathbb P$. A basic result says that if $E$ is an e-value, then $1/E$ is a p-value. This is easy to see using Markov's inequality and \eqref{eq:eval}:
$$
    \mathbb P(1/E\leq \alpha)=\mathbb P(E\geq 1/\alpha)\leq\alpha,
$$
where we used the fact that $\mathbb E_\mathbb P(E)\leq 1$ for all $\mathbb P\in\mathcal P_0$.

\subsubsection{Combination}\label{sec:combine}
We show that it is easy to combine e-values, even with unknown dependence. Assume that we have $T$ e-values, $E_1,\dots,E_T$ where $E_t$ was computed from $\mathcal G^{(t)}$, and assume that we do not know the dependence between $\mathcal G^{(t)}$ and $\mathcal G^{(t')}$ for $t'\neq t$. Thus, we also do not know the dependence between $E_t$ and $E_{t'}$ for $t\neq t'$. A natural, and in some sense ``best'' approach to combine e-values is with the arithmetic mean, i.e.,
$$
    \mathbb M(e_1,\dots,e_T)=\frac1T\sum_{t=1}^T e_t.
$$
It is easy to see that the arithmetic mean of e-values is still an e-value as, for any $\mathbb P\in \mathcal P_0$,
$$
  \mathbb E_\mathbb P\{\mathbb M(E_1,\dots,E_T)\}
  =\frac1T\sum_{t=1}^T \mathbb E_\mathbb P(E_t)
  \leq \frac1T\sum_{t=1}^T 1
  =1.
$$

We refer the interested reader to \cite{vovk2021values, ramdas2024hypothesis} for more details on combining e-values. Intuitively, if e-values from the individual snapshots are large, then the arithmetic mean will also be large. In other words, if there is strong evidence for community structure on individual snapshots, then there will be strong evidence for community structure on the temporal network.

\section{Methodology}\label{sec:method}

\subsection{Proposed test}\label{sec:prop}
We propose the following hypothesis test for community structure in temporal networks.
\begin{enumerate}
    \item For $t\in\{1,\dots,T\}$, compute a p-value, $P_t$, on $\mathcal G^{(t)}$.
    \item Convert each p-value to an e-value, $E_t$.
    \item Find the average of the e-values, $\bar E_T=\frac1T\sum_{t=1}^T E_t$.
    \item Reject $H_0$ if $\bar E_T$ is ``large.''
\end{enumerate}

\paragraph{Step 1: Compute p-values: }First, we must compute a p-value on each graph snapshot $\mathcal G^{(t)}$ for $t\in\{1,\dots,T\}$. This p-value corresponds to the evidence for/against a community structure on the static component $\mathcal G^{(t)}$. We stress that any static hypothesis test for community structure can be used as long as it yields a valid p-value. We primarily use the approach from \cite{bickel2016:aa}, with details provided in the Supplementary Materials.

\paragraph{Step 2: Convert p-values to e-values:} Since $\mathcal G^{(t)}$ is likely dependent on $\mathcal G^{(t')}$ for $t'\neq t$, the corresponding p-values $P_t$ and $P_{t'}$ will also be dependent. In general, it is quite difficult to combine dependent p-values, especially if this dependence is unknown. To circumvent this issue, we convert each p-value to an e-value since the latter are trivial to combine. To do this, we use a {\it (p-e) calibrator}. A calibrator is a non-negative decreasing function $g:[0,\infty)\to [0,\infty]$ with integral at most one such that $g(x)=0$ for all $x\in(1,\infty)$ and $g(P)$ is an e-value if $P$ is a p-value. In other words, the calibrator transforms a p-value into an e-value; the reverse transformation is also possible, but we will not need this in the present paper. Some examples of calibrators are given in \cite{shafer.vovk.martingale} and  \cite{vovk2021values}. Specifically, for any $\kappa\in(0,1)$, 
\[ g_\kappa(p) = \kappa \, p^{\kappa - 1} \]
is a calibrator.  In case one does not want to specify the parameter $\kappa$, there are two natural options.  First, one might choose $\kappa$ to maximize $g_\kappa(p)$ at each $p$, leading to 
\[ g_\text{max}(p) = \begin{cases}
        -\exp(-1)/(p\log p)&p\leq \exp(-1),\\
        1&\text{otherwise},
    \end{cases}
\] 
but this is not a calibrator. We refer to this as {\it max} in the rest of the paper.  Second, and less greedy, one might average over $\kappa$ to get 
\[ g_\text{avg}(p) = \frac{1-p+p\log p}{p(-\log p)^2}, \]
the calibrator we call {\it avg}. We consider each of these three approaches in our simulations.


\paragraph{Step 3: Average e-values:} Once the p-values have been converted to e-values, we combine them by taking the arithmetic mean. We showed in Section \ref{sec:back} that not only does this yield a valid e-value, but it can also be shown that this is the best way to combine them with unknown dependence \citep{vovk2021values}. More generally, for any positive weights $w_1,\dots,w_T$ such that $\sum_{t=1}^T w_t=1$,
$$
    \tilde E_T
    =\sum_{t=1}^T w_t E_t
$$
is also a valid e-value. Of course, the user would then need to choose the weights, but, e.g., selecting weights such that $w_1<w_2<\dots<w_T$ would place a greater importance on more recent snapshots, which would be reasonable if it was plausible that a community structure was ``evolving'' in the time index $t$. Finally, if we knew that the snapshots $\mathcal G^{(t)}$ were independent, then we could also multiply the e-values to form a new e-value, but in general we do not expect the network realizations to be independent.

\paragraph{Step 4: Rejection decision:} After averaging the e-values, we are left with a single e-value which quantifies the evidence for or against the presence of a community structure in the temporal network. Typically in hypothesis testing, the user chooses some threshold $\alpha\in(0,1)$ as their type I error rate, and then rejects if the p-value is smaller than $\alpha$ with common choices $\alpha=0.05$ or $\alpha=0.01$. On the other hand, for e-values, we could reject the null hypothesis when it is ``large.'' While there is no universally agreed upon threshold, $E>20$ yields a sensible rejection threshold, yielding a type I error of no more than $\alpha=0.05$ \citep{wang2023tiny}. That being said, we do not necessarily endorse the use of 20 as {\it the} threshold for decision making. Instead, the raw e-value should be considered as the strength of evidence against the null hypothesis of ``no community structure,'' and the practitioner can choose the appropriate level, albeit greater than 1.

\subsection{Properties}
We first state the key property of the previous test with the following theorem.\\

\noindent
{\sc Theorem 1.} {\it Let $\mathcal G^{(1)},\dots,\mathcal G^{(T)}$ be a sequence of snapshots from a temporal network and let $\bar E_T$ be constructed as in Section \ref{sec:prop}. Then $\bar E_T$ is an e-value in the sense of \eqref{eq:pval}.}
\\

Theorem 1 follows directly from the basic definition of e-values described in Section \ref{sec:evals} and allows $\bar E_T$ to be used for hypothesis testing. The novelty here lies in the application to testing on temporal networks. Indeed, to the authors' knowledge, this is the first hypothesis test for community structure in temporal networks.

While not explicitly stated, the proposed test implicitly adopts its null hypothesis of ``no community structure'' via the static community detection hypothesis test. Many methods use the ER model, e.g., \cite{bickel2016:aa, yuan2022testing, yanchenko2024generalized}. Here, the probability of an edge between any two nodes is independently $p$, i.e., $A_{ij}\mid p\stackrel{\text{iid.}}{\sim}\mathsf{Bernoulli}(p)$. We write $\bA\sim \mathbb P_{ER}(p)$ as short-hand to denote the ER data-generating distribution with edge probability $p$. Then $\mathcal M_1$ as the set of all possible network generating matrices, and $\mathcal P_0=\{\mathbb P_{ER}(p)\mid p\in(0,1)\}$. Tautologically, this model does not encode any community structure. Furthermore, certain data-generating models implicitly encode the ER model as a null model \citep[e.g.,][]{sasahara2021social}. 

Practitioners may be wary of this choice of null model. In particular, the model may be so simple that virtually all real-world networks diverge from it, whether they possess a community structure or not. In such cases, the configuration model may be a more appropriate null model \citep{lancichinetti2010statistical, palowitch2017significance}. For ease of exposition, we focus on the ER null model and \cite{bickel2016:aa} hypothesis testing framework, but we stress that the method can easily be generalized to other null models and/or hypothesis tests. Indeed, any community structure hypothesis test for static networks can be adapted to this framework. We explore such generalizations in the real-data analysis.

As for the alternative hypothesis, this is generally not explicitly encoded in the hypothesis test. Instead, the definition of ``community structure'' will be determined by the user depending on their data-generating model. A sensible choice would be the stochastic block model (SBM) \citep{holland83} but it can be more general. As our preferred hypothesis test from \cite{bickel2016:aa} does not encode a specific alternative hypothesis, the proposed test does not make any modeling assumptions of the network-generating process. This means that the test can be used for a wide range of models and settings, e.g., the community structure changes over time as nodes change groups, new communities form and/or disappear, new nodes are added to the network, etc. Similarly, we did not need to specify the dependence between successive observations $\mathcal G^{(t)}$ and $\mathcal G^{(t+1)}$ because the averaging of e-values is still a valid e-value, regardless of their dependence. 

Lastly, we stress that the main reason for leveraging an e-value framework is that it easily allows for combining evidence across different snapshots (here, via averaging). If we had used p-values directly, then it would be highly non-trivial to combine results, even if we assume a data-generating model. That being said, by averaging e-values, the proposed approach implicitly tests for an ``average'' community structure over the temporal evolution of the network. This means, for example, that if the snapshot ordering was shuffled, the proposed approach would still yield the same e-value. We do not consider this a feature or a bug of the method, but rather a reality that users need to be aware of. Indeed, as we discuss further in the Conclusion, defining community structure on a temporal network is quite difficult, and since, to our knowledge, there are no other existing hypothesis tests for this situation, we view this as a sensible first solution. Of course, more explicitly accounting for the evolution in the network could yield a more powerful test. We leave this investigation for future work. Our method does, however, implicitly account for correlation in the network, e.g., if every snapshot was identical (perfect correlation), then the e-value would not be an average and instead would be the value from only a single snapshot.

\section{Simulation Study}\label{sec:sim}

\subsection{Set-up}
Here we perform numerical simulations to study the properties of the proposed hypothesis test, demonstrating its ability to combine results across dependent network snapshots. Thus, we assume that the number of observations, $T$, was fixed before the experiment began.

As the test is not tied to a particular data-generating model, we consider three different network-generating models. Since we use the static test from \cite{bickel2016:aa}, for each model, the hypotheses being tested are
$$
    H_0:\text{Erdos-Renyi model vs. }H_1:\text{not }H_0.
$$
We vary parameter settings which control the strength of the community structure or the number of nodes, and report the median e-value over 100 Monte Carlo (MC) iterations. To clarify, the individual e-values are calculated as the average e-value from each snapshot, while the median is taken over the MC replicates. We compare the proposed test using five different calibrators: {\it max}, {\it avg} and $\kappa=0.25,0.50,0.75$. Recall that {\it max} is not a proper calibrator but we may refer to it as one for ease of exposition. We prefer the median as opposed to the mean as an infinite e-value from just a single MC iteration can obscure the overall trends. While we are primarily interested in reporting the e-values of the test directly, we also report the type I error and power of the test for a rejection threshold of $\bar E_T>20$ in the Supplemental Materials, in addition to variability results. We stress that he goal of these simulations is not to identify the specific communities, but rather to test for the presence/absence of a community structure. Finally, code to implement the proposed test and reproduce the simulation studies is available from the first author's GitHub page: \url{https://github.com/eyanchenko/tempComDet#}.

\subsection{Correlated SBM}\label{sec:corSBM}

\subsubsection{Model}
First, we propose a novel model to generate correlated observations with a stochastic block model \citep{holland83}. The stochastic block model (SBM) is a popular model to generate networks with a community structure in static networks. We present the basic ideas before generalizing to temporal networks. Given some network with $n$ nodes and $K$ communities, let $\bc\in\{1,\dots,K\}^n$ be such that $c_i=k$ if node $i$ is in community $k$. Here, $\bc$ represents the latent community structure. 
Given $\bc$, the probability of an edge between nodes $i$ and $j$, $P_{ij}$, is $P_{ij}=B_{c_i,c_j}$, where ${\bf B}\in (0,1)^{K\times K}$ is the block-probability matrix such that $B_{k,k'}$ is the probability of an edge between nodes in community $k$ and $k'$ for $k,k'\in\{1,\dots,K\}$. If $B_{k,k'}=b$ for all $k,k'\in\{1,\dots,K\}$, then this model reduces to the ER (null) model. On the other hand, if the diagonal entries of ${\bf B}$ are larger than those on the off-diagonal, then the SBM generates networks with assortative community structure (alternative hypothesis).

To extend this to the temporal setting, we are inspired by the graph matching literature \citep[see, e.g.][]{lyzinski14a} to generate a (potentially correlated) sequence of network realizations, or snapshots. At time step, $t=1$, we generate a static SBM network, i.e., 
$$
    A_{ij}^{(1)}|\bc,\bB\stackrel{\text{ind.}}{\sim}\mathsf{Bernoulli}(B_{c_i,c_j})\text{ for all }i,\ j.
$$
Then for all $i,j$ and $t\geq 2$,
\begin{equation}\label{eq:sbm}
    A_{ij}^{(t)}|A_{ij}^{(t-1)},\bc,\bB
    \stackrel{ind.}{\sim} 
    \begin{cases}
        \mathsf{Bernoulli}\{B_{c_i,c_j}+\rho(1-B_{c_i,c_j})\}&\text{if }A_{ij}^{(t-1)}=1\\
        \mathsf{Bernoulli}\{B_{c_i,c_j}(1-\rho)\}&\text{if }A_{ij}^{(t-1)}=0
    \end{cases}
\end{equation}
This formulation ensures that $\mathsf{Cor}(A_{ij}^{(t)},A_{ij}^{(t-1)})=\rho$, which we prove in the Supplemental Materials. As special cases, if $\rho=0$, then the snapshots are independent, and if $\rho=1$, then they are identical. This resembles a first-order Markov process as the next snapshot only depends on the current one, and $\rho$ controls the amount of correlation between the successive realizations. Additionally, this model assumes that the latent community labels, $\bc$, do not vary with time, but we relax this assumption in the subsequent model.

\subsubsection{Settings}
\begin{figure}[p]
    \centering
    \begin{subfigure}[b]{0.7\textwidth}
        \centering
        \includegraphics[width=\textwidth]{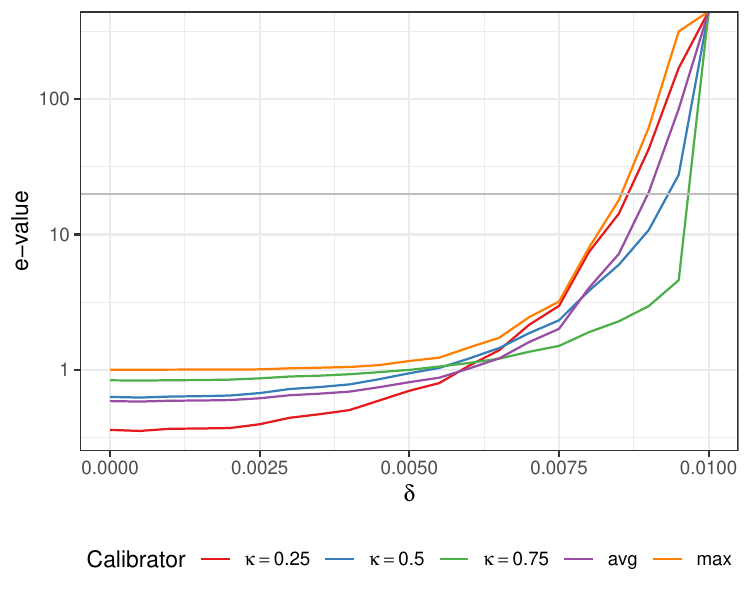}
        \caption{Varying $\delta$.}
        \label{fig:1a}
    \end{subfigure}

    \begin{subfigure}[b]{0.7\textwidth}
        \centering
        \includegraphics[width=\textwidth]{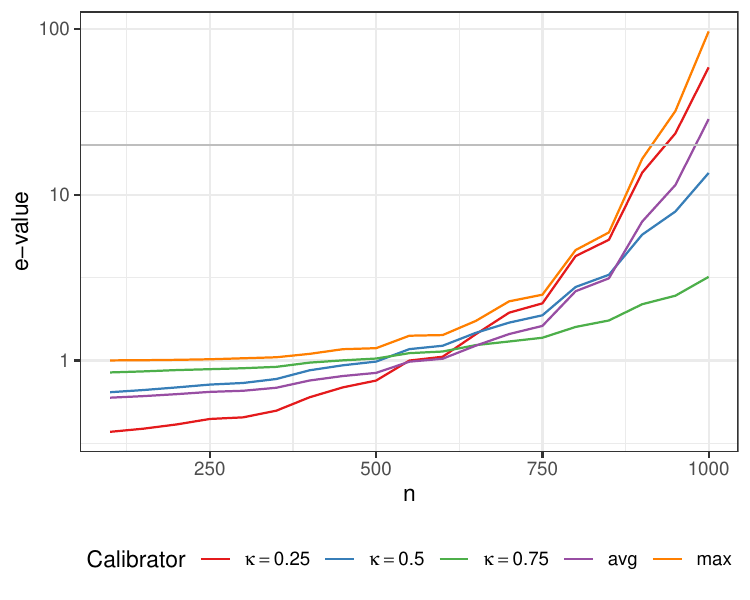}
        \caption{Varying $n$.}
        \label{fig:1c}
    \end{subfigure}
    
    \caption{Median e-value over 100 MC simulations for correlated SBM networks with $\rho=0.25$. Grey line at $E=20$ corresponding to $\alpha=0.05$ rejection threshold.}
    \label{fig:corSBM}
\end{figure}

To generate networks using this model, we set $n_1=\cdots=n_T\equiv n=1000$ where $T=10$ and assume $K=2$ such that
$$
    \bB
    =
    \begin{pmatrix}
        b+\delta/2&b-\delta/2\\
        b-\delta/2&b+\delta/2
    \end{pmatrix}
    .
$$
Then we set $b=0.01$ such that $\delta>0$ corresponds to the strength of the community structure in the network. We assume that $\bc$ is fixed for all $t$ and that 80\% of the nodes are in group 1 and the remaining are in group 2. With $\rho=0.25$, we generate temporal networks using \eqref{eq:sbm}. In Figure \ref{fig:1a}, we vary $\delta\in\{0,5\times10^{-4},\dots,0.01\}$, i.e., increasing community structure.  In Figure \ref{fig:1c}, we fix $\delta=0.009$ and vary $n\in\{100, 150, \dots, 1000\}$, i.e., increasing network size. The results are similar for the same settings but with $\rho=0.75$, so we leave these to the Supplemental Materials.

\subsubsection{Results}
For $\delta>0$, the alternative hypothesis is true so we expect to see large e-values. Indeed, as $\delta$ increases, the median e-value monotonically increases for each calibrator. The {\it max} and $\kappa=0.25$ calibrators consistently have slightly larger e-values for $\delta>0.006$ with $\kappa=0.75$ having the lowest. For the increasing $n$ setting, $\delta>0$ so we expect large e-values. All calibrators, save $\kappa=0.75$, yield large values for $n=1000$, and the relative performance of the calibrators is similar to that of the previous setting. These results show that the proposed testing method can detect community structure in SBMs, with the {\it max} and $\kappa=0.25$ calibrators performing the best, even when the networks (and therefore the e-values) are correlated. Moreover, the test yields a high power when the community structure is relatively weak, e.g., $\delta=0.01$ implies that an intra-community edge is only 1\% more likely than an inter-community edge.  

\subsection{Dynamic SBM}\label{sec:dynSBM}
\subsubsection{Model}
While the previous model allows for correlation between successive realizations of the network, it does not allow the node communities to change with time. To allow for this property, we discuss the dynamic SBM \citep{matias2017statistical}. Recall that we represent our temporal network as a sequence of adjacency matrices $\bA^{(1)},\dots,\bA^{(T)}$. Additionally, we have community labels $\bc^{(1)},\dots,\bc^{(T)}$ such that $\bc^{(t)}\in \{1,2,\dots,K\}^{n}$ and $c_i^{(t)}=k$ means that node $i$ is in group $k$ at time $t$. Thus, the community labels are now allowed to vary with time. We also define $\bc_i=(c_i^{(1)},\dots,c_i^{(T)})^\top$ to represent the group membership for node $i$ over time.

\cite{matias2017statistical} assume that $\bc_1,\dots,\bc_n$ are independent and identically distributed random variables. For a given node, however, $\bc_i$ is a Markov chain defined by transition matrix $\bpi\in(0,1)^{K\times K}$ and initial stationary distribution $\boldsymbol{\alpha}=(\alpha_1,\dots,\alpha_K)^\top$. In words, if node $i$ is in community $k$ at time $t$, then the probability that it is in group $k'$ at time $t+1$ is $\pi_{k,k+1}$ for $k,k'\in\{1,\dots,K\}$. Given $\bc^{(1)},\dots,\bc^{(T)}$, we generate a SBM as before:
\begin{equation}\label{eq:dynSBM}
    A_{ij}^{(t)}|\bc_i^{(t)},\bc_j^{(t)},\bB\sim\mathsf{Bernoulli}(B_{c_i^{(t)},c_j^{(t)}})
\end{equation}
where $\bB\in(0,1)^{K\times K}$ is the block model probability matrix. The authors show that, for identifiability, the community labels and block-model parameters cannot both vary with time, so we consider $\bB$ as fixed over time. This also means that if all the entries of $\bB$ are the same, then we recover the ER null model again. Lastly, notice that conditional on the group labels, the consecutive snapshots are independent, unlike the previous model.

\subsubsection{Settings}
\begin{figure}[p]
    \centering
    \begin{subfigure}[b]{0.7\textwidth}
        \centering
        \includegraphics[width=\textwidth]{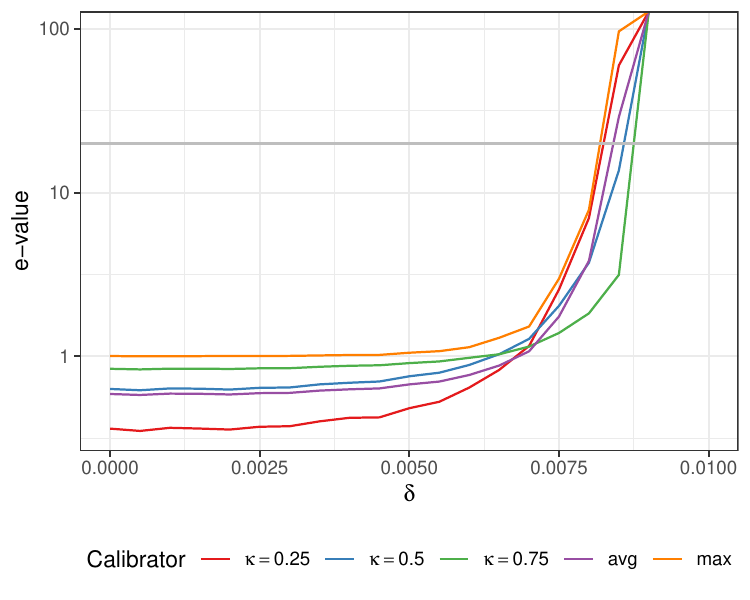}
        \caption{Varying $\delta$.}
        \label{fig:2a}
    \end{subfigure}
    \hfill
    
    \begin{subfigure}[b]{0.7\textwidth}
        \centering
        \includegraphics[width=\textwidth]{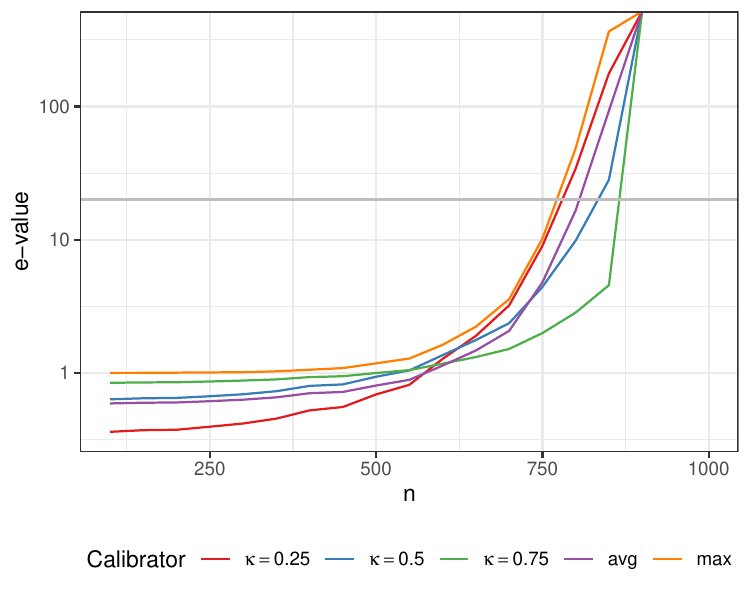}
        \caption{Varying $n$.}
        \label{fig:2c}
    \end{subfigure}
    
    \caption{Median e-value over 100 MC simulations for dynamic SBM networks with $\bpi_1$. Grey line at $E=20$ corresponding to $\alpha=0.05$ rejection threshold.}
    \label{fig:dynSBM}
\end{figure}

We again let $n=1000$, $T=10$, $K=2$ and take
$$
    \bB
    =
    \begin{pmatrix}
        b+\delta/2&b-\delta/2\\
        b-\delta/2&b+\delta/2
    \end{pmatrix}
    ,
$$
with $b=0.01$ and $\boldsymbol{\alpha}=(0.80, 0.20)^\top$. We set
$$
    \bpi_1
    =
    \begin{pmatrix}
        0.90&0.10\\
        0.10&0.90
    \end{pmatrix}
$$
which discourages nodes from switching groups, vary $\delta\in\{0,5\times10^{-4},\dots,0.01\}$ and generate networks using \eqref{eq:dynSBM}. The results are in Figure \ref{fig:2a}. We also fix $\delta=0.009$ and vary $n\in\{100, 150,\dots,1000\}$ in Figure \ref{fig:2c}. The same settings but with 
$$
    \bpi_2
    =
    \begin{pmatrix}
        0.60&0.40\\
        0.40&0.60
    \end{pmatrix}
$$
were also considered and these results are in the Supplemental Materials.

\subsubsection{Results}

When $\delta>0$, the alternative hypothesis is true so we expect to see large e-values. The median e-value is low for all methods until around $\delta=0.007$, at which point they all begin to increase. Again, the {\it max} calibrator consistently has the largest e-values. For $\delta\geq0.009$, all p-values were 0 leading to an infinite e-value, and is therefore not plotted. The trends are similar for increasing $n$; for $n\geq900$, all e-values were infinite and not plotted. These results show that the proposed testing method can detect community structure even when it is weak and the community labels vary temporally.

\subsection{Dynamic DCBM}\label{sec:dynDCBM}
\subsubsection{Model}
For the final simulation, we consider a temporal degree-corrected block model (DCBM). The DCBM was first proposed in \cite{karrer2011stochastic} to model the fact that many real-world networks have degree heterogeneity. As before, let $\bc$ correspond to the community labels and $\bB$ represent the community edge probabilities. We also introduce node weights, $\btheta=(\theta_1,\dots,\theta_n)^\top$ such that if $\theta_i$ is large, then we expect node $i$ to have relatively more edges.
Then the probability distribution of the network is defined as
\begin{equation}\label{eq:dynDCBM}
    A_{ij}\mid \bc,\bB,\btheta\stackrel{\text{iid.}}{\sim}\mathsf{Bernoulli}(\theta_i\theta_j B_{c_i,c_j}).
\end{equation}
Thus, the probability of an edge between nodes $i$ and $j$ depends not only on their community assignment (through $\bB$) but also their relative importance (via $\theta$). Note that if $\theta_i\equiv\theta$ for all $i$, then we recover the non-degree-corrected SBM. Furthermore, if $B_{k,k'}=b$ for all $k,k'$, then the DCBM reduces to the Chung-Lu (CL) model \citep{chung2002average} which does not possess community structure, but also differs from the ER model, i.e., $P_{ij}\propto\theta_i\theta_j$.

There are various ways for this model to vary temporally including through the community assignments $(\bc)$, block probabilities $(\bB)$ or weight parameters $(\btheta)$ \citep{wilson2019modeling}. For simplicity, we follow the approach of \cite{matias2017statistical} to let the community labels change with time, in addition to varying $\theta_i$ \citep{wilson2019modeling}. Specifically, we generate
$$
    \theta_i^{(t)}\stackrel{\text{iid.}}{\sim}\mathsf{Uniform}(1-\varepsilon/2,1+\varepsilon/2)
$$
and $\bc^{(t)}$ as in Section \ref{sec:dynSBM}. Then conditional on $\bc^{(t)},\btheta^{(t)}$ and $\bB$, we generate independent snapshots using \eqref{eq:dynDCBM}.

\subsubsection{Settings}
We use similar settings as in Section \ref{sec:dynSBM}. Let $n=1000$, $T=10$, $K=2$ and take $\bB$ as before with $b=0.01$ and $\boldsymbol{\alpha}=(0.80, 0.20)^\top$. We use $\bpi_1$ to model the community transition probabilities and take $\varepsilon=0.6$. We vary $\delta\in\{0,5\times10^{-4},\dots,0.01\}$, i.e., increasing community structure, and generate networks using \eqref{eq:dynDCBM}. The results are in Figure \ref{fig:3a}. Next, we fix $\delta=0.009$ and vary $n\in\{100, 150,\dots,1000\}$ with results in Figure \ref{fig:3c}). We repeat these settings with $\varepsilon=0.2$ (decreased degree heterogeneity) in the Supplemental Materials.

\subsubsection{Results}

\begin{figure}[p]
    \centering
    \begin{subfigure}[b]{0.7\textwidth}
        \centering
        \includegraphics[width=\textwidth]{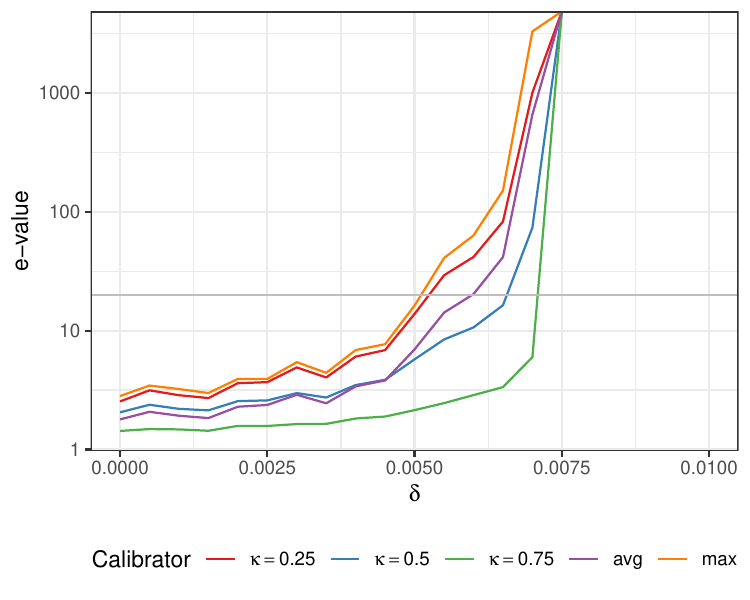}
        \caption{Varying $\delta$.}
        \label{fig:3a}
    \end{subfigure}
    \hfill
    
    \begin{subfigure}[b]{0.7\textwidth}
        \centering
        \includegraphics[width=\textwidth]{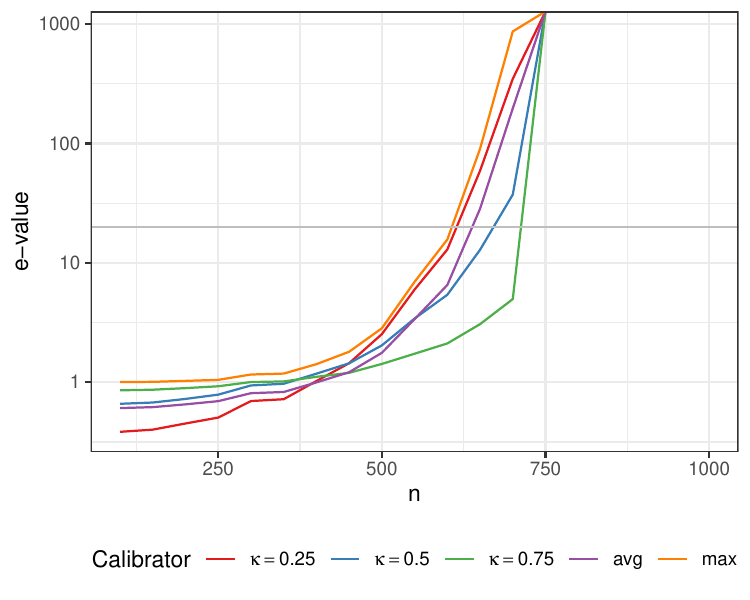}
        \caption{Varying $n$.}
        \label{fig:3c}
    \end{subfigure}
    
    \caption{Median e-value over 100 MC simulations for dynamic DCBM networks with $\varepsilon=0.6$. Grey line at $E=20$ corresponding to $\alpha=0.05$ rejection threshold.}
    \label{fig:dynDCBM}
\end{figure}

The results are similar to the previous ones. As $\delta$ and $n$ increase, the median e-value also increases. When $\delta$ is varied, the {\it max} and $\kappa=0.25$ calibrators yield the largest e-values for all values of $\delta$. Again, the $\kappa=0.75$ calibrator tends to have the smallest e-value. The main difference from the previous settings is that the e-value increase occurs for smaller $\delta$ and $n$. This is because the added  heterogeneity from the $\btheta$ parameters means the model diverges from the ER model more quickly than a basic SBM. Indeed, in Figure \ref{fig:3a}, the median e-values are greater than 1 even when $\delta=0$ (no block model structure). Even though $\delta=0$ and the null hypothesis appears to be true, the DCBM behaves like a CL model which still differs from the ER model, causing the slight inflation in the e-value. This is a key limitation of the method from \cite{bickel2016:aa} in that it only tests against the ER null. As seen here, it is possible for a network to diverge from the ER null without possessing a community structure. This idea will be more fully explored in the real-data analysis.

\section{Real-data analysis}\label{sec:real}

\subsection{Choice of null model}
We turn our attention to studying real-world networks, further demonstrating the ability of the proposed test to combine information across dependent network realizations. In the previous sections, we leveraged the static community structure hypothesis test from \cite{bickel2016:aa}. The proposed e-value testing procedure, however, works more generally for any test that yields p-values on static networks. Indeed, in many cases, the ER model may not be a satisfactory null model as it is too easy to reject. In other words, almost all real-world networks diverge from a simple ER model, even if they do not have a community structure. Specifically, in the dynamic DCBM simulations, we found that even if there was no block community structure $(\delta=0)$, the CL model alone led to slightly larger e-values then we would have expected under the null (ER model). Moreover, in the following examples, we found that each network (save {\it Kenya}) yielded an infinite e-value when using the method from \cite{bickel2016:aa}. If every network yields an infinite e-value, then it may not be useful metric for discriminating community strength in networks.

In light of these observations, we apply the bootstrap testing approach from \cite{yanchenko2024generalized} in order to leverage the Chung-Lu (CL) model as the null model, i.e., testing $H_0:\text{Chung-Lu model}$ versus $H_1:\text{not }H_0$. We prefer this null as it can model degree heterogeneity while still not exhibiting community structure, making it a more flexible null model. Indeed, the CL model is a close cousin of the configuration model which is commonly used as a null model in community detection algorithms \citep[e.g.,][]{newman2006modularity}.

\subsection{Bootstrap method}
We briefly describe the bootstrap testing method from \cite{yanchenko2024generalized}. The authors begin by defining the expected edge density difference (E2D2) parameter to quantify the strength of a community structure. Specifically, for adjacency matrix $\bA$ and community labels $\bc$, the E2D2 statistic is defined as scaled difference between the intra- and inter-community edge density, i.e.,
\begin{equation}
    U(\bA,\bc)
    =\frac1K\frac{\hat p_{in}(\bc)-\hat p_{out}(\bc)}{\hat p},
\end{equation}
where
$$
    {\bar p_{\text{in}}}(\bc) 
    = \frac{1}{\sum_{k=1}^K\binom{n_k}{2}}\sum_{i<j} P_{ij}\ \mathbb I(c_i=c_j)  \ \ \ \text{ and }\ \ \ 
    \bar p_{\text{out}}(\bc) 
    = \frac{1}{\sum_{k> l} n_k n_l}\sum_{i<j} P_{ij}\ \mathbb I(c_i\neq c_j) ,
$$
and where $\mathbb I(\cdot)$ is the indicator function and $n_k$ is the number of nodes in community $k$, i.e., $n_k=\sum_{i=1}^n\mathbb I(c_i=k)$ for $k\in\{1,\dots,K\}$. The authors also propose a greedy algorithm to approximate the maximum value, i.e.,
\begin{equation}\label{eq:e2d2}
    \tilde U(\bA)
    =\max_{\bc}\{U(\bA,\bc)\}.    
\end{equation}

The bootstrap test proceeds as follows. Given the observed adjacency matrix $\bA$, $\tilde U(\bA)$ is calculated as in \eqref{eq:e2d2}. Since we are testing against the null hypothesis of a CL model, we need to generate CL networks that resemble $\bA$. To do this, we first estimate the weight vector $\btheta$ with using the rank 1 adjacency spectral embedding (ASE) \citep{sussman2012consistent}:
$$
    \hat\btheta
    =|\lambda_1|^{1/2} \, {\bf u}
$$
where $\lambda_1$ is the largest eigenvalue of $\bA$ (in magnitude) and ${\bf u}$ is the corresponding eigenvector. Then for the $b$th bootstrap iteration, we sample from $\hat\btheta=(\hat\theta_1,\dots,\hat\theta_n)^\top$ with replacement to obtain $\btheta^{(b)}=(\theta_1^{(b)},\dots,\theta_n^{(b)})^\top$ and then sample a network $\bA^{(b)}$ using $\btheta^{(b)}$. We compute $\tilde U(\bA^{(b)})=\max_{\bc}\{U(\bA^{(b)},\bc)\}$ and repeat this process for $B$ iterations. Finally, the p-value is 
$$
    \frac1B\sum_{b=1}^B\mathbb I\{\tilde U(\bA^{(b)})\geq\tilde U(\bA)\}.
$$
See \cite{yanchenko2024generalized} for further details and properties of this test. We briefly note that only the largest connected component of the graph is used when computing $\tilde U(\bA)$ and the number of communities $K$ is found using the Fast Greedy Algorithm, with an upper-bound set at $\tilde n^{1/2}$, where $\tilde n$ is the number of nodes in the largest connected component. In this real-data analysis, we use this test to compute the p-value on each network snapshot, and then flip to e-values using the procedure described in Section \ref{sec:prop}.

\subsection{Data sets}
To study the performance of the proposed hypothesis test under various settings, we consider five real-world networks of various sizes and domains. The first four networks are all proximity networks which record an edge if two people are close to each other: {\it Kenya} \citep{kiti2016quantifying}, {\it Reality} \citep{eagle2006reality}, {\it Hospital} \citep{Vanhems:2013} and {\it High School 1} \citep{mastrandrea2015contact}. The remaining network is constructed from online communication on a social media platform: {\it College} \citep{panzarasa2009patterns}. For each network, we apply the bootstrap hypothesis test with CL null from \cite{yanchenko2024generalized} with $B=1,000$ bootstrap iterations. We also look at the effect of the number of snapshots by varying  $T\in\{5,10\}$, where we again assume that this choice was made before the data were collected. We emphasize that the temporal network is given with each edge having a time stamp. Thus, the total observation time of the network process is fixed, but we can vary the number of snapshots based on how we bin the data. Indeed, increasing $T$ should not be confused with increasing the length of the observation period in this setting.

\subsection{Results}
For each network, the number of nodes $n$ and e-value are reported in Table \ref{tab:real_data}. We use the $\kappa=0.25$ calibrator because it yielded the largest e-values in the simulation studies when $E>1$, and a rejection threshold of $\bar E_T>20$. {\it Kenya} does not show evidence of a statistically significant community structure given the small e-values. {\it Hospital} and {\it College} have slightly larger e-values, but still not large enough to warrant rejection of the null hypothesis. For {\it Hospital}, this finding accords with that in \cite{yanchenko2024generalized}, which found no community structure compared to the CL null for this network, though this was when the network was treated as static, i.e., $T=1$. For {\it Reality} and {\it High School 1}, however, there is very strong evidence to reject the null hypothesis and claim that these networks exhibit a statistically meaningful community structure. In general, the results do not seem to be overly sensitive to the choice of $T$. Regardless, we stress that in practice this choice should be carried out in a meaningful way based on the domain.

\begin{table}[]
    \centering
    \begin{tabular}{lc|cc}
    \multicolumn{1}{l}{ }&
    \multicolumn{1}{c|}{ }&
    \multicolumn{2}{c}{$T$} \\
        Network & $n$ & 5 & 10   \\\hline
        Kenya & 52 & 0.4 & 0.4  \\
        Reality & 64 & $\infty$ & $\infty$\\
        Hospital & 75  & 9.4 & 6.5 \\ 
        High School 1 & 312 & $\infty$ & $\infty$ \\
        College & 1899 & 5.0 & 5.7
    \end{tabular}
    \caption{Real-world network e-values using the Chung-Lu null model, bootstrap hypothesis test and $\kappa=0.25$ calibrator.}
    \label{tab:real_data}
\end{table}

\section{Conclusion}\label{sec:disc}
In this work, we proposed a novel test for community structure in temporal networks which, to the knowledge of the authors, is the first of its kind. The proposed test finds the p-value from a static hypothesis test on each snapshot before converting these to e-values and averaging. A large e-value gives evidence in favor of community structure being present in the temporal network. Moreover, the e-value framework easily accommodates arbitrary dependence in the network. The simulation studies showed that our test can accurately detect the presence/absence of community structure under a range of data-generating models. 

The results of the real-data analysis elucidate some of the challenges of temporal network testing, as well as the limitations of the proposed approach. For example, consider the Reality network where  we found that the e-values were infinite. This resulted from the last snapshot yielding a p-value of 0, implying an infinite e-value. Of course, averaging any set of numbers where one is infinite will also be infinite. In other words, if the p-value is extremely small (or 0) on just a single snapshot, then the e-value will explode. Indeed, in the Reality network example, if we remove the last snapshot from the averaging, then the e-value is only 1.2 and 0.4 for $T=5,10$, respectively, indicating no significant community structure. On the other hand, the e-value was large for almost all snapshots in the High School 1 network results.

These examples serve as a case-study in the difficulties of testing for community structure in temporal networks. In this work, the approach to average e-values implicitly assumes that the ``strength'' of the community structure is roughly constant with time. In other words, we do not expect there to be extremely strong community structure in one snapshot, but none later in the observation process. This implicit assumption also arises because by averaging, we believe that subsequent temporal observations are yielding more evidence about some underlying community structure. The dynamic SBM/DCBM examples show that our framework does allow for community memberships to change, but the underling strength is roughly constant. Specifically, in all simulations, we assumed that $\bB$, the block probabilities, were constant with time. 

This raises an important question which we have hitherto not explicitly addressed: what does it mean for a temporal network to have community structure? This question is discussed in \cite{cazabet2023challenges}. Perhaps the simplest definition is {\it fixed} community structure where the labels and underlying strength of the community structure are invariant with time. The correlated SBM discussed in Section \ref{sec:corSBM} would be one example of this. Next, we may have {\it persistent} community structure where the community labels are now allowed to change over time, but the strength of the community structure is still roughly constant. The dynamic SBM (Section \ref{sec:dynSBM}) and dynamic DCBM (Section \ref{sec:dynDCBM}) both exhibit this kind of community structure, as did the {\it High School 1} network. The proposed hypothesis testing formulation is sensible for either of these cases.

On the other hand, the community structure may change dramatically over the life of the network. For example, new communities may appear (Birth), disappear (Death), join together (Merge), separate (Split), disappear and then reappear (Resurgence) or do something else entirely. These concepts are most connected with the community labels, but the edge probabilities may also vary time. The {\it Reality} network seemed to exhibit at least one of these feature as it evolved between exhibiting and not exhibiting community structure, what we might refer to as {\it intermittent} community structure. As a result, the e-value based test was extremely sensitive to these fluctuations. In this case, it is not clear whether the proposed test is even meaningful.

Another related possibility is {\it accumulating} community structure where the strength of the community structure increases with time. An example of this case comes from models like that of \cite{sasahara2021social}. In this work, the authors propose a model for echo chambers formation in online social networks, describing how starting from a relatively well-mixed population can quickly lead to strong communities if edges are formed and removed based on some simple heuristics. In this case, the community structure is, by design, not constant, and only arises after a certain point. In this case, we may be more interested in change point detection to know {\it when} a community structure arises \citep[e.g.,][]{wilson2019modeling}. 

Regardless, a precise characterization of temporal network community structure is needed to develop a sensible hypothesis test. For example, is the intermittent community structure like that of {\it Reality} a meaningful structure? How ``intermittent'' can it be until it is no longer community structure? Does it make sense to test for community structure where communities are coming, going, merging and splitting through the process? In this work, we proposed a statistically valid test by extending the definition from static networks, but a more nuanced definition may be needed depending on the particular behavior of the network of interest. These and similar questions will also need to be addressed if the methods in this work are extended to test for other network properties, e.g., core-periphery structure \citep{yanchenko2023core}. 

As another possible direction for future work, our proposed framework can easily be extended to other situations where multiple realizations of the network arise, i.e., {\it multilayer networks.} \citep{kivela2014multilayer}. For example, there could be a network where nodes are researchers, and one layer of the network represents collaborations as edges, a different layer uses edges to denote citations, and a third layer corresponds to a friendship network. If the practitioner is interested in testing for the significance of community structure across these different layers, then our approach could easily be applied.

On the theoretical and methodological side, one possible improvement would be to construct the network-specific e-values directly, rather than applying a calibrator to a known p-value.  The intuition would go roughly as follows.  The individual p-values are themselves relatively efficient, and the ``dream-world'' e-value would be simply the reciprocal of that efficient p-value.  Unfortunately, the reciprocal of a p-value is not an e-value, and, roughly speaking, the calibrator strategically inflates the p-value by a small amount before taking the reciprocal and returning an e-value.  It is this preliminary inflation that, while validity is preserved, leads to a less-efficient-than-necessary e-value.  A direct e-value construction like in \citet{grunwald2024safe, numeraire} would generally be more efficient, and the construction of such an e-value is the focus of future research.

\bibliographystyle{apalike}
\bibliography{refs}

\clearpage

\begin{center}
    Supplemental Materials
\end{center}

\section*{Static hypothesis test}
We describe a hypothesis test for community structure on static networks from \cite{bickel2016:aa}.
Let $\mathcal G\equiv \mathcal G^{(1)}$ be a static network and $\bA\equiv \bA^{(1)}$ the associated adjacency matrix and define the Erdos-Renyi model as the null model. Specifically, define $\hat p$ as the estimated probability of an edge between any two nodes and let
$$
    \hat p
    =\frac1{n(n-1)}\sum_{i=1}^n\sum_{j=1}^n A_{ij}.
$$
Furthermore, let $\hat{\bf P}$ be the estimated network-generating matrix for an ER model where $\hat P_{ij}=\hat p$ for all $i\neq j$ and $\hat P_{ii}=0$ for all $i$. We define $\tilde\bA$ as a ``standardized'' version of the adjacency matrix, i.e.,
$$
    \tilde \bA = \frac{1}{\sqrt{(n-1)\hat p(1-\hat p)}}(\bA-\hat{\bf P}).
$$
If the null hypothesis is true that the network was generated from an ER model, then \cite{bickel2016:aa} show that the largest eigenvalue $\lambda_1(\tilde\bA)$ (considering negative signs), asymptotically follows the Tracy-Widom distribution with index 1, i.e.,
$$
    \lambda_1(\tilde \bA)\sim TW_1.
$$
This null distribution can be used to compute the p-value for the hypothesis test. 

The authors show that the convergence to the Tracy-Widom distribution is slow, so they propose a bootstrap correction for small $n$. The idea is to generate ER networks with $\hat p$, find the largest eigenvalue on these networks, and then appropriately shift and scale the original result. The details are laid out in Algorithm \ref{alg:hyp} where we let $\mu_{TW}$ and $\sigma_{TW}$ be the theoretical mean and standard deviation of a Tracy-Widom distribution with index 1. We use this algorithm unless otherwise noted.

\begin{algorithm}
\SetAlgoLined
\KwResult{p-value}
 {\bf Input: } adjacency matrix $\bA$; \;

$\hat p = \sum_{i=1}^n\sum_{j=1}^n A_{ij}/\{n(n-1)\}$\;

$\gamma = \frac{1}{\sqrt{(n-1)\hat p(1-\hat p)}}\lambda_1(\bA-\hat{\bf P})$\;

\For{$b=1:50$}{
    $\tilde\bA_b\sim\mathsf{ER}(n,\hat p)$\;

    $\tilde \gamma_b = \frac{1}{\sqrt{(n-1)\hat p(1-\hat p)}}\lambda_1(\tilde\bA_b-\hat{\bf P})$\;
}

$\hat\mu_\gamma = \text{mean}(\tilde\gamma_b)$\;

$\hat\sigma_\gamma = \text{stdev}(\tilde\gamma_b)$\;

$\gamma' = \mu_{TW} + \frac{\gamma-\hat\mu_\gamma}{\hat\sigma_\gamma}\sigma_{TW}$\;

p-value = $\mathbb P(X>\gamma')$ where $X\sim TW_1$\;

 \Return{p-value}
 
 \caption{Bootstrap correction to static hypothesis test for community structure.}
 \label{alg:hyp}
\end{algorithm}

\section*{Proofs}
Note that we can drop the subscripts on $A$ and we let $A_t=A^{(t)}$.\\
{\it Lemma 1. For all $t$, $A_t\sim\mathsf{Bernoulli}(p)$ marginally.}\\ 
{\it Proof.} We show this by induction. The claim trivially holds for $t=1$. For $t=2$, we have
$$
    p(a_2|a_1)
    =\{p+\rho(1-p)\}^{a_1a_2}\{(1-p)(1-\rho)\}^{a_1(1-a_2)}\{p(1-\rho)\}^{(1-a_1)a_2}\{1-p(1-\rho)\}^{(1-a_1)(1-a_2)},\ a_2\in\{0,1\}
$$
Thus,
\begin{multline*}
    p(a_1,a_2)
    =\\
    \{p+\rho(1-p)\}^{a_1a_2}
    \{(1-p)(1-\rho)\}^{a_1(1-a_2)}
    \{p(1-\rho)\}^{(1-a_1)a_2}
    \{1-p(1-\rho)\}^{(1-a_1)(1-a_2)}
    p^{a_1}(1-p)^{1-a_1},\ a_1,a_2\in\{0,1\}.
\end{multline*}
and
\begin{multline*}
    p(a_2)
    =\sum_{a_1=0}^1 p(a_1,a_2)
    = 
    (1-p)\{p(1-\rho)\}^{a_2}
    \{1-p(1-\rho)\}^{(1-a_2)}
    +
    p\{p+\rho(1-p)\}^{a_2}
    \{(1-p)(1-\rho)\}^{(1-a_2)}\\
    =p^{a_2}(1-p)^{1-a_2}
\end{multline*}
In other words, $A_2\sim\mathsf{Bernoulli}(p)$ marginally. For the induction step, assume that $A_t\sim\mathsf{Bernoulli}(p)$ and we want to show that $A_{t+1}$ is also Bernoulli distributed with success probability p. It is easy to see that if we repeat the above work replacing $a_1,a_2$ with $a_t,a_{t+1}$, respectively, then the same result holds, completing the proof of the lemma.\\

\noindent
{\it Claim.} $\mathsf{Cor}(A_{ij}^{(t+1)}, A_{ij}^{(t)})=\rho$.\\
{\it Proof.} We again proceed by induction.
For $t=1$, we have
$$
    \mathbb E(A_1A_2)
    =\sum_{a_1=0}^1\sum_{a_2=0}^1 a_1a_2p(a_1,a_2)
    =p(1,1)
    =p\{p+\rho(1-p)\},
$$
$$
    \mathbb E(A_1)=\mathbb E(A_2)=p
$$
Thus,
$$
    \mathsf{Cov}(A_1,A_2)
    =p\{p+\rho(1-p)\} - p^2
    = \rho p(1-p)
$$
Additionally, $\mathsf{Var}(A_1)=\mathsf{Var}(A_2)=p(1-p)$ so
$$
    \mathsf{Cor}(A_1,A_2)
    =\rho
$$
as we hoped to show.

Now we show the induction step. Assume the claim holds for time $t$ and we want to show that it holds for time $t+1$. Based on Lemma 1, $A_t$ and $A_{t+1}$ are both marginally Bernoulli distributed, so we can repeat the above work replacing $A_1,A_2$ with $A_t,A_{t+1}$ and the result directly follows. $\square$

\section*{Simulation study type I error and power results}
\begin{figure}[htbp]
    \centering
    \begin{subfigure}[b]{0.75\textwidth}
        \centering
        \includegraphics[width=\textwidth]{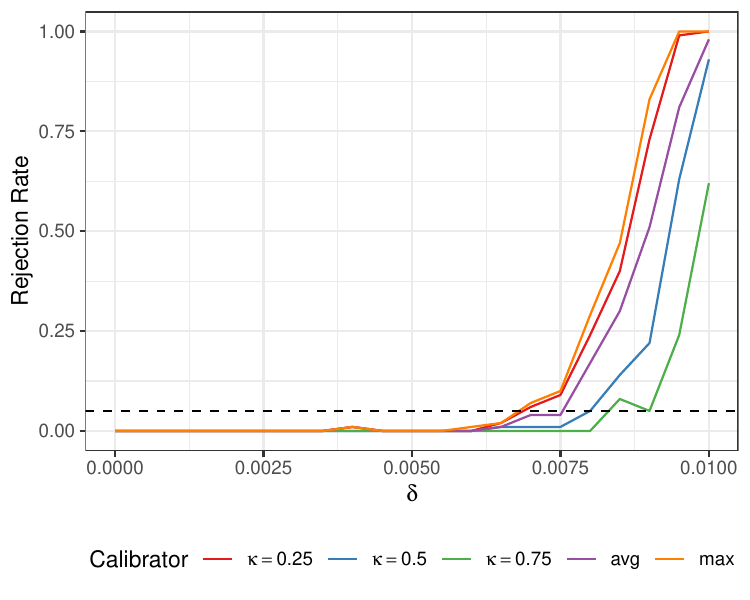}
        \caption{Varying $\delta$}
        \label{fig:1b_supp}
    \end{subfigure}
    \begin{subfigure}[b]{0.75\textwidth}
        \centering
        \includegraphics[width=\textwidth]{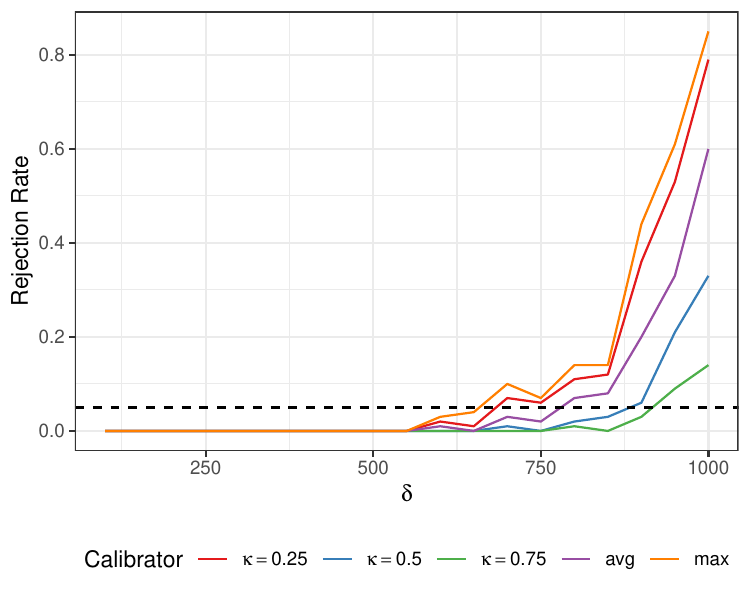}
        \caption{Varying $n$}
        \label{fig:1d_supp}
    \end{subfigure}
    
    \caption{Simulation rejection rates for correlated SBM networks with $\rho=0.25$ and rejection threshold $\bar E_T>20$. Dashed black line corresponds to $\alpha=0.05$.}
    \label{fig:corSBM_power}
\end{figure}

\begin{figure}[htbp]
    \centering
    \begin{subfigure}[b]{0.75\textwidth}
        \centering
        \includegraphics[width=\textwidth]{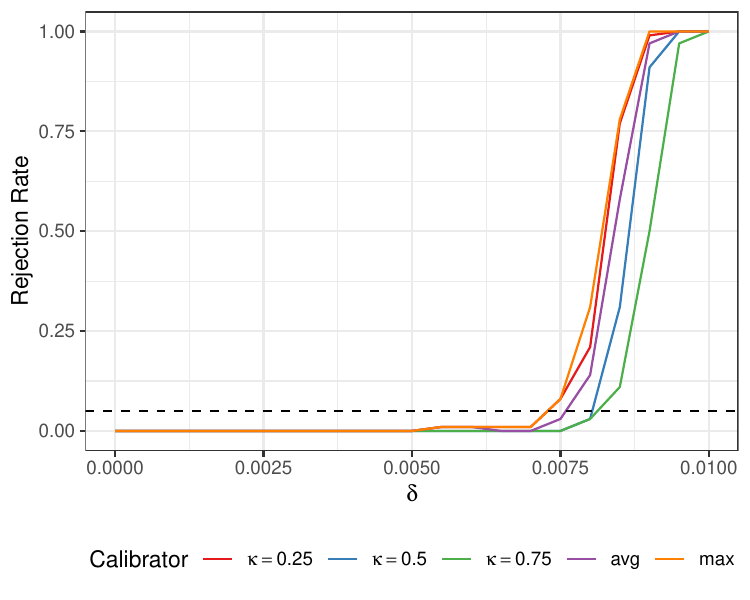}
        \caption{Varying $\delta$}
        \label{fig:1b_supp}
    \end{subfigure}
    \begin{subfigure}[b]{0.75\textwidth}
        \centering
        \includegraphics[width=\textwidth]{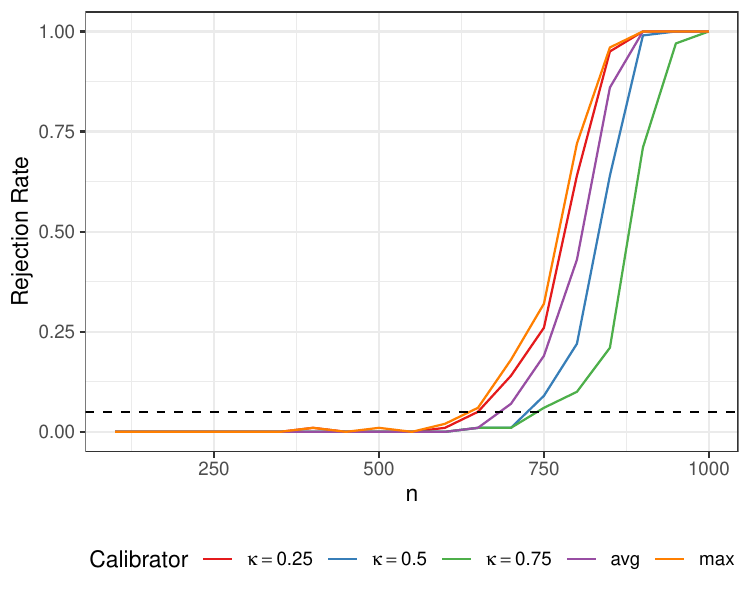}
        \caption{Varying $n$}
        \label{fig:1d_supp}
    \end{subfigure}
    \caption{Simulation rejection rates for dynamic SBM networks with $\boldsymbol{\pi}_1$ and rejection threshold $\bar E_T>20$. Dashed black line corresponds to $\alpha=0.05$.}
    \label{fig:dynSBM_power}
\end{figure}

\begin{figure}[htbp]
    \centering
    \begin{subfigure}[b]{0.75\textwidth}
        \centering
        \includegraphics[width=\textwidth]{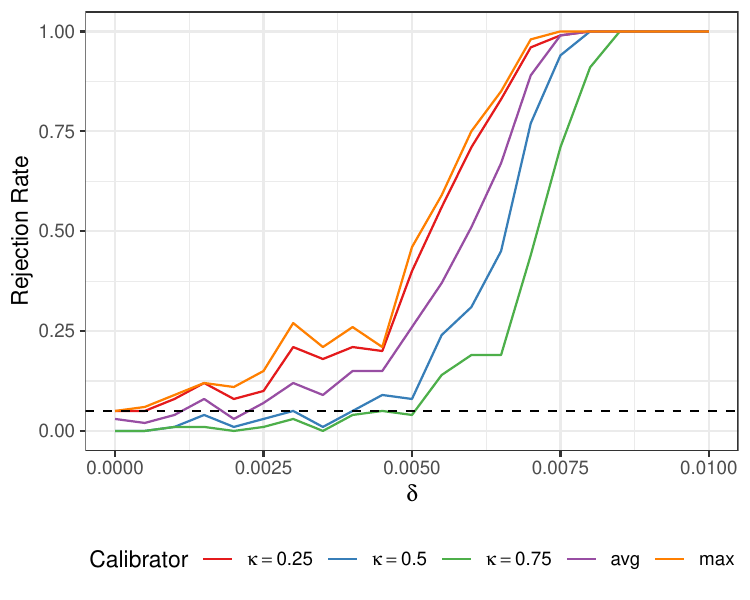}
        \caption{Varying $\delta$}
        \label{fig:1b_supp}
    \end{subfigure}
    \begin{subfigure}[b]{0.75\textwidth}
        \centering
        \includegraphics[width=\textwidth]{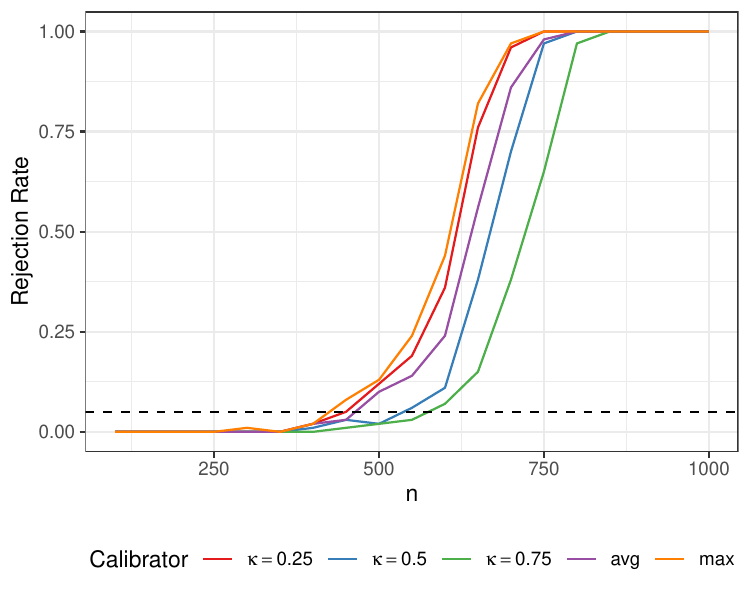}
        \caption{Varying $n$}
        \label{fig:1d_supp}
    \end{subfigure}
    \caption{Simulation rejection rates for dynamic DCBM networks with $\epsilon=0.6$ and rejection threshold $\bar E_T>20$. Dashed black line corresponds to $\alpha=0.05$.}
    \label{fig:dynDCBM_power}
\end{figure}

In the main manuscript, we are primarily interested in reporting the raw e-value as a continuous measure of evidence for/against the null hypothesis. If a typical rejection decision is required, however, a binary threshold can be used. Towards this end, we report the rejection rate for the settings in the main manuscript using $\bar E_T>20$ as the rejection decision, corresponding to a type I error of no more than $\alpha=0.05$. The results are in Figures \ref{fig:corSBM_power},\ref{fig:dynSBM_power},\ref{fig:dynDCBM_power}. When $\delta=0$, the null hypothesis is true and the test maintains a low rejection rate. For $\delta>0$, however, the alternative hypothesis is true and the test yields a large rejection rate.

\section*{Simulation study variability}

\begin{figure}[htbp]
    \centering
    \begin{subfigure}[b]{0.75\textwidth}
        \centering
        \includegraphics[width=\textwidth]{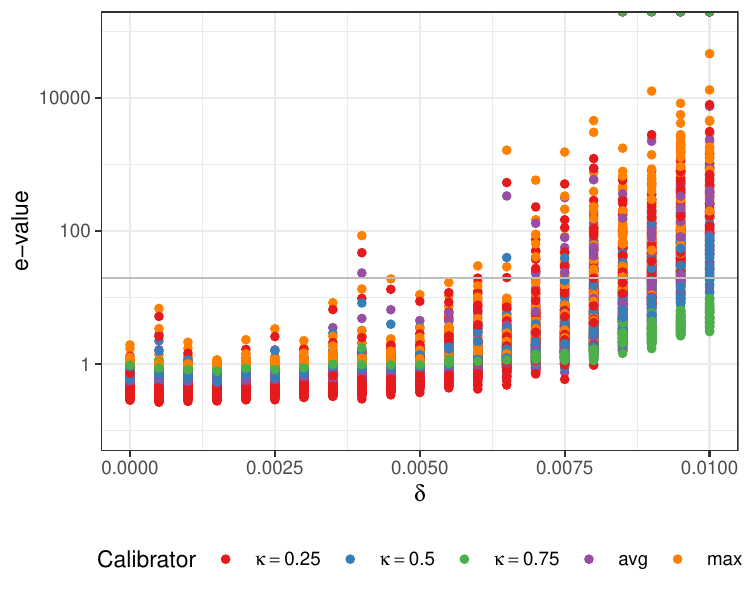}
        \caption{Varying $\delta$}
        \label{fig:1b_supp}
    \end{subfigure}
    \begin{subfigure}[b]{0.75\textwidth}
        \centering
        \includegraphics[width=\textwidth]{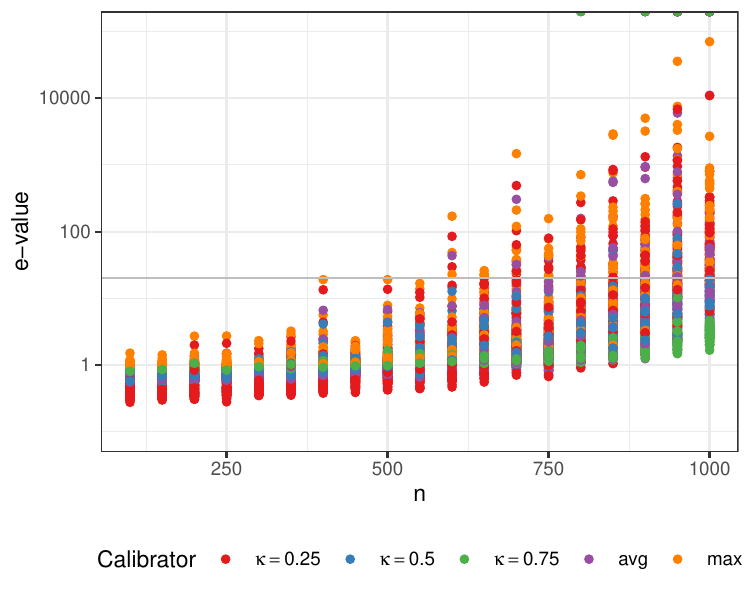}
        \caption{Varying $n$}
        \label{fig:1d_supp}
    \end{subfigure}
    
    \caption{Simulation variability results for correlated SBM networks with $\rho=0.25$. Grey line at $E=20$ corresponding to $\alpha=0.05$ rejection threshold.}
    \label{fig:corSBM_var}
\end{figure}

In the main manuscript, we report the median e-value for each calibrator. Here, we plot the actual value for each MC iteration to give a sense of the measure of variability. In Figure \ref{fig:corSBM_var}, we plot the results for the correlated SBM setting for increasing $\delta$ and $n$ with $\rho=0.25$. Given that this is on the log-scale, there is clearly a large degree of variability in the e-values across all settings, particularly for larger $\delta$ and $n$.

\section*{Additional simulation results}

\begin{figure}[htbp]
    \centering
    \begin{subfigure}[b]{0.75\textwidth}
        \centering
        \includegraphics[width=\textwidth]{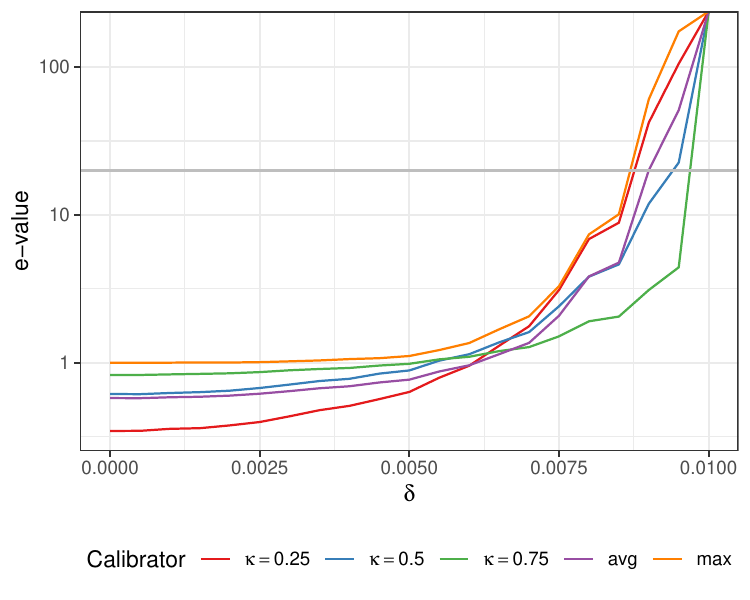}
        \caption{Varying $\delta$}
        \label{fig:1b_supp}
    \end{subfigure}
    \begin{subfigure}[b]{0.75\textwidth}
        \centering
        \includegraphics[width=\textwidth]{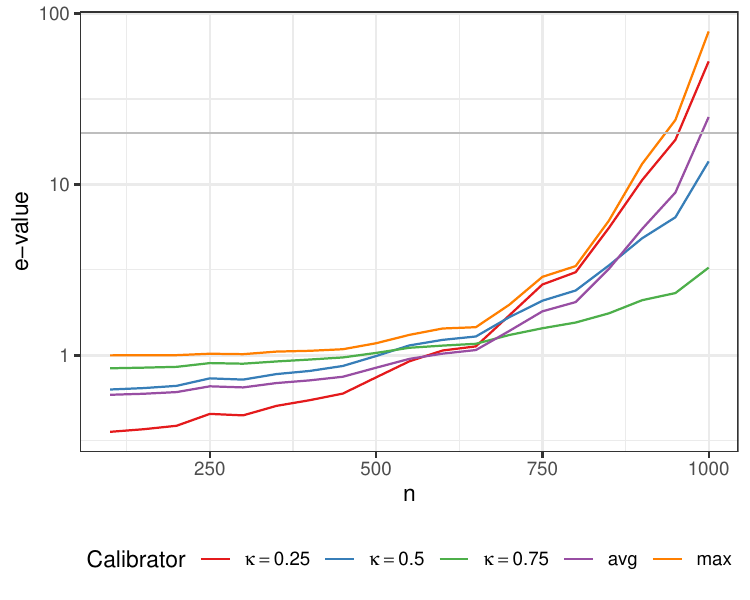}
        \caption{Varying $n$}
        \label{fig:1d_supp}
    \end{subfigure}
    
    \caption{Median e-value over 100 MC simulations for correlated SBM networks with $\rho=0.75$. Grey line at $E=20$ corresponding to $\alpha=0.05$ rejection threshold.}
    \label{fig:corSBM_supp}
\end{figure}

\begin{figure}[htbp]
    \centering
    \begin{subfigure}[b]{0.75\textwidth}
        \centering
        \includegraphics[width=\textwidth]{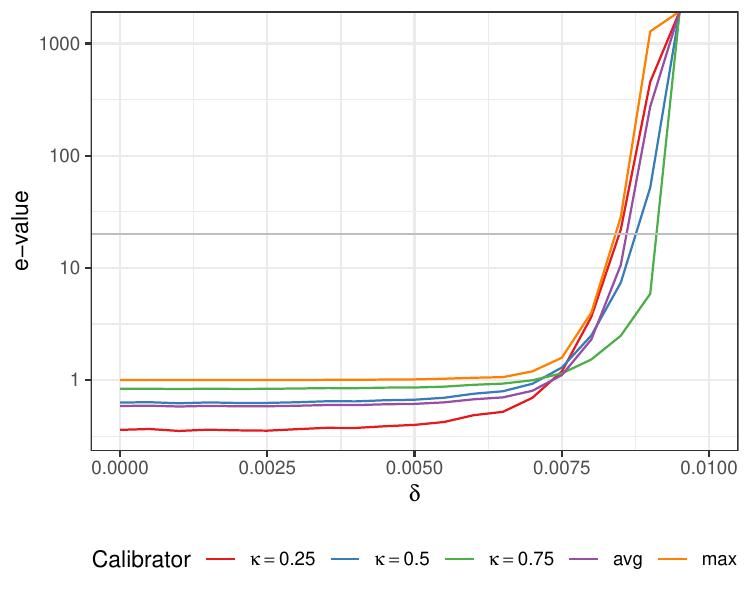}
        \caption{Varying $\delta$}
        \label{fig:2b}
    \end{subfigure}

    \begin{subfigure}[b]{0.75\textwidth}
        \centering
        \includegraphics[width=\textwidth]{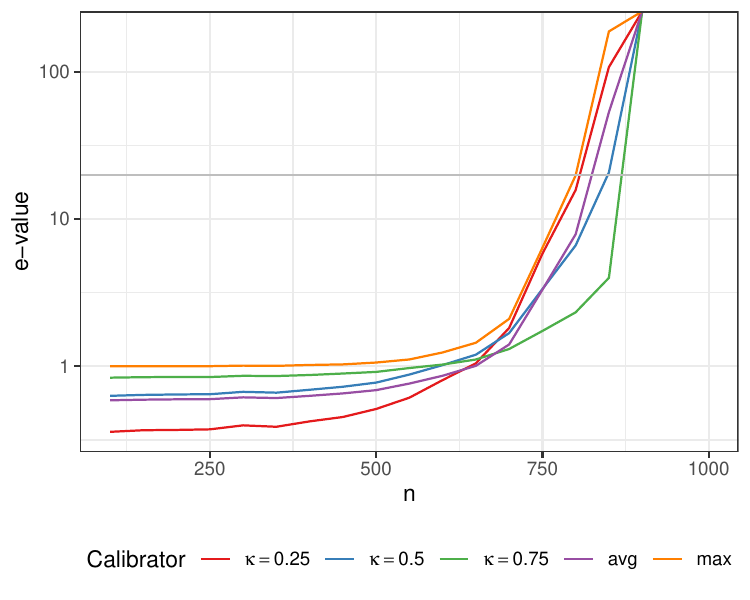}
        \caption{Varying $n$}
        \label{fig:2d}
    \end{subfigure}
    
    \caption{Median e-value over 100 MC simulations for dynamic SBM networks with $\bpi_2$. Grey line at $E=20$ corresponding to $\alpha=0.05$ rejection threshold.}
    \label{fig:dynSBM_supp}
\end{figure}

\begin{figure}[htbp]
    \centering
    \begin{subfigure}[b]{0.75\textwidth}
        \centering
        \includegraphics[width=\textwidth]{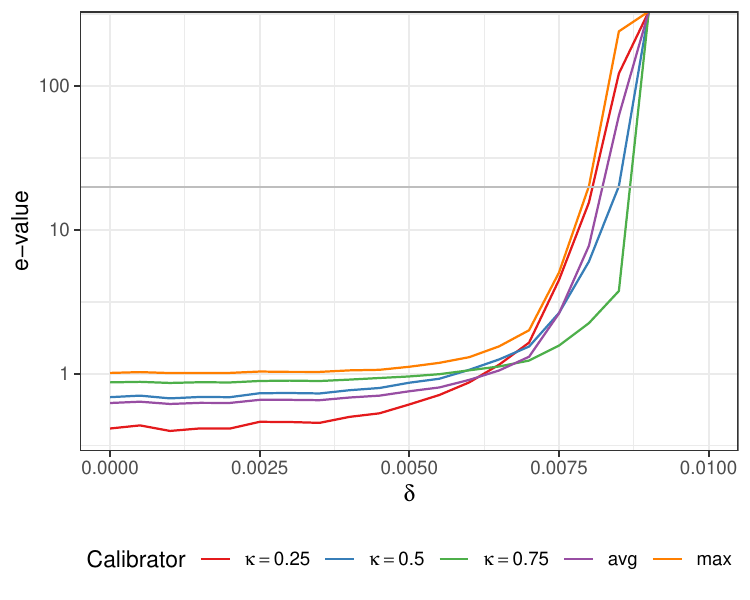}
        \caption{Varying $\delta$}
        \label{fig:3b}
    \end{subfigure}

    \begin{subfigure}[b]{0.75\textwidth}
        \centering
        \includegraphics[width=\textwidth]{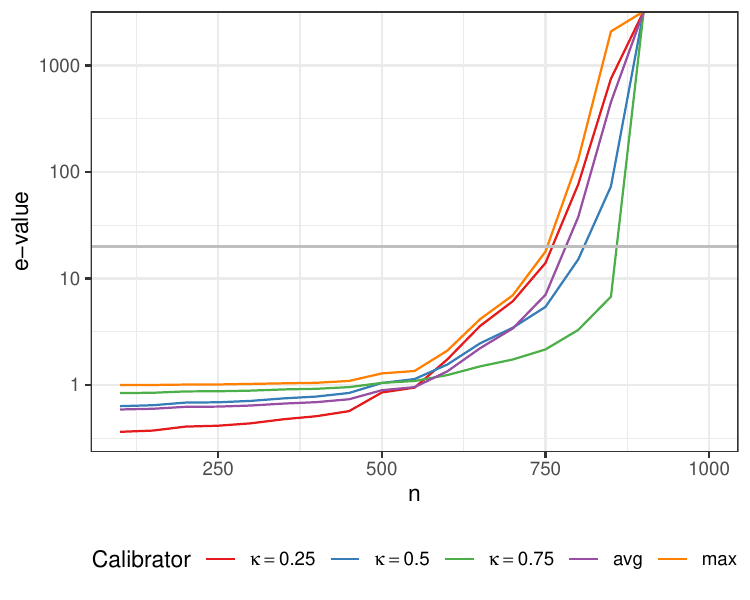}
        \caption{Varying $n$}
        \label{fig:3d}
    \end{subfigure}
    
    \caption{Median e-value over 100 MC simulations for dynamic DCBM networks with $\varepsilon=0.2$. Grey line at $E=20$ corresponding to $\alpha=0.05$ rejection threshold.}
    \label{fig:dynDCBM_supp}
\end{figure}

We include additional simulation results for the correlated SBM, dynamic SBM and dynamic DCBM in Figures \ref{fig:corSBM_supp}, \ref{fig:dynSBM_supp}, \ref{fig:dynDCBM_supp}, respectively. The trends are similar to those in the main manuscript.

\end{document}